\documentclass{article}

\usepackage{microtype}
\usepackage{graphicx}
\usepackage{subcaption}
\usepackage{booktabs} % for professional tables

\usepackage{enumitem}
\usepackage{microtype}
\usepackage{xspace}
\usepackage{multirow}
\usepackage{seqsplit}
\usepackage{xcolor}
\usepackage{hyperref}
\usepackage[ruled,vlined]{algorithm2e}
\hypersetup{
    colorlinks=true,
    urlcolor=blue,
    linkcolor=blue,
    citecolor=blue
}
\newcommand{\sys}{\textsc{OServe}\xspace}
\usepackage{xcolor}
\newcommand{\switch}{ad hoc model switching\xspace}
\definecolor{lightgreen}{RGB}{144,238,144}
\definecolor{darkcyan}{HTML}{008B8B}
\usepackage{hyperref}
\usepackage{xurl}

\usepackage[accepted]{icml2026}

\usepackage{amsmath}
\usepackage{amssymb}
\usepackage{mathtools}
\usepackage{amsthm}

\usepackage[capitalize,noabbrev]{cleveref}

\theoremstyle{plain}

\theoremstyle{definition}

\theoremstyle{remark}

\usepackage[textsize=tiny]{todonotes}

\icmltitlerunning{\sys: Accelerating LLM Serving via Spatial-Temporal Workload Orchestration}

\begin{document}

\twocolumn[
  \icmltitle{\sys: Accelerating LLM Serving via Spatial-Temporal \\ Workload Orchestration}

  % It is OKAY to include author information, even for blind submissions: the
  % style file will automatically remove it for you unless you've provided
  % the [accepted] option to the icml2026 package.

  % List of affiliations: The first argument should be a (short) identifier you
  % will use later to specify author affiliations Academic affiliations
  % should list Department, University, City, Region, Country Industry
  % affiliations should list Company, City, Region, Country

  % You can specify symbols, otherwise they are numbered in order. Ideally, you
  % should not use this facility. Affiliations will be numbered in order of
  % appearance and this is the preferred way.
  \icmlsetsymbol{equal}{*}

  \begin{icmlauthorlist}
    \icmlauthor{Youhe Jiang}{equal,yyy}
    \icmlauthor{Fangcheng Fu}{equal,comp}
    \icmlauthor{Taiyi Wang}{equal,yyy}
    \icmlauthor{Guoliang He}{yyy}
    \icmlauthor{Eiko Yoneki}{yyy}
    %\icmlauthor{}{sch}
    %\icmlauthor{}{sch}
  \end{icmlauthorlist}

  \icmlaffiliation{yyy}{Department of Computer Science, University of Cambridge, Cambridgeshire, UK}
  \icmlaffiliation{comp}{School of Artificial Intelligence, Shanghai Jiao Tong University, Shanghai, China}
  \icmlcorrespondingauthor{Eiko Yoneki}{eiko.yoneki@cl.cam.ac.uk}

  % You may provide any keywords that you find helpful for describing your
  % paper; these are used to populate the "keywords" metadata in the PDF but
  % will not be shown in the document
  \icmlkeywords{Machine Learning, ICML}

  \vskip 0.3in
]

% this must go after the closing bracket ] following \twocolumn[ ...

% This command actually creates the footnote in the first column listing the
% affiliations and the copyright notice. The command takes one argument, which
% is text to display at the start of the footnote. The \icmlEqualContribution
% command is standard text for equal contribution. Remove it (just {}) if you
% do not need this facility.

% Use ONE of the following lines. DO NOT remove the command.
% If you have no special notice, KEEP empty braces:
\printAffiliationsAndNotice{\icmlEqualContribution}  % no special notice (required even if empty)
% Or, if applicable, use the standard equal contribution text:
% \printAffiliationsAndNotice{\icmlEqualContribution}

\begin{abstract}
Serving Large Language Models (LLMs) can benefit immensely from parallelizing both the model and input requests across multiple devices, but incoming workloads exhibit substantial \textit{spatial} and \textit{temporal} heterogeneity. Spatially, workloads comprise heterogeneous requests with varying compute and memory demands. Temporally, workload composition varies over time. Nevertheless, existing systems typically assume spatially uniform and temporally stable workloads, employing a homogeneous, static model deployment. This mismatch between the assumption and real-world spatial-temporal heterogeneity results in suboptimal performance. We present \sys, an LLM serving system with \textit{heterogeneous} and \textit{flexible} model deployment that addresses both spatial and temporal heterogeneity. First, \sys introduces a novel \textit{workload-aware scheduling algorithm} that optimizes heterogeneous model deployments according to real-time workload characteristics. Second, \sys proposes an efficient \textit{workload-adaptive switching method} that migrates model deployments in response to predicted workload changes. Experiments on real-world traces show that \sys improves performance by up to 2$\times$ (average: 1.5$\times$) compared to state-of-the-art serving systems.
\end{abstract}

\section{Introduction}

Large Language Models (LLMs) such as OPT~\cite{zhang2022opt}, Llama~\cite{dubey2024llama}, gpt-oss~\cite{agarwal2025gpt}, Gemini~\cite{reid2024gemini}, Claude~\cite{claude3}, and Mixtral~\cite{jiang2024mixtral} have demonstrated exceptional performance across a range of advanced applications~\cite{peng2023study,jeon2023large,copilot}. 
To democratize LLMs, it has become a timely and important topic to optimize the efficiency of LLM serving. 

With LLMs deployed across increasingly diverse applications, inference workloads exhibit substantial heterogeneity along two key dimensions: (\textbf{\underline{i}}) \textbf{Spatial heterogeneity:} Workloads comprise heterogeneous requests with significant variance in resource demands across concurrent requests—certain requests are compute-intensive (e.g., chat and summarization tasks with short output lengths), while others are memory-intensive (e.g., generation and transformation tasks with long output lengths)~\cite{bai2025longbench,gao2024omni,zhao2024blendserve,agrawal2024taming,jiang2025demystifying}. (\textbf{\underline{ii}}) \textbf{Temporal heterogeneity:} Request composition and arrival rates fluctuate dynamically over time in response to evolving user behavior and application usage patterns~\cite{patel2024splitwise,jaiswal2025serving}.

% \begin{figure}[!t]
%     \centering
%     \includegraphics[width=\linewidth]{imgs/combined_figure.pdf}
%     \vspace{-1em}
%     \caption{The first row shows performance comparisons between data, tensor, and pipeline parallelism (DP, TP, PP) across different resource allocations and workload types. The four workload types come from real-world traces in the Azure Public Dataset~\cite{azuredataset}, detailed in \S\ref{sec:workload predictor}. The second row compares model parallelism strategies with 8 GPUs, where (n$_1$, n$_2$) represents TP and PP degrees. Long input and output requests have sequence lengths exceeding 1024 and 64.}
%     \vspace{-1em}
%     \label{fig:tpt}
% \end{figure}

To meet the substantial computational and memory requirements, LLMs are commonly deployed in a distributed manner using parallelism strategies~\cite{li2024llm,miao2025towards}, including data parallelism (model replication)~\cite{li2023alpaserve}, tensor parallelism~\cite{shoeybi2019megatron}, and pipeline parallelism~\cite{huang2019gpipe}. However, these strategies exhibit distinct trade-offs when serving workloads with spatial and temporal heterogeneity.

\begin{figure}
    \centering
    \includegraphics[width=\linewidth]{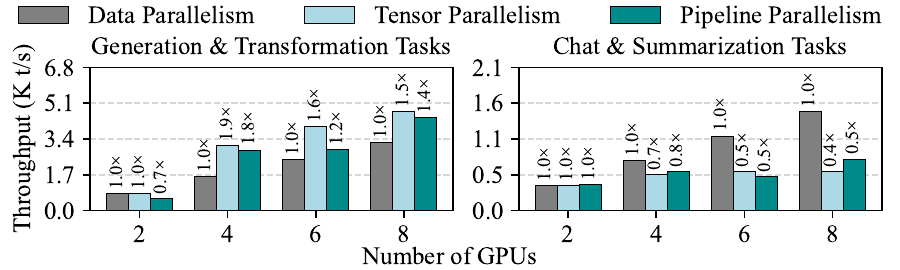}
    \caption{Performance comparisons of different parallelism strategies across resource allocations and workload types. The two workload types are subsampled from real-world traces in the Azure Public Dataset~\cite{patel2024splitwise}.}
    \label{fig:spatial}
\end{figure}

\begin{figure}
    \centering
    \includegraphics[width=\linewidth]{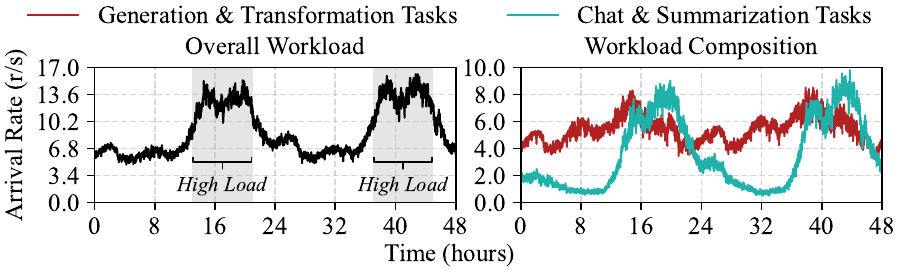}
    \caption{Temporal evolution of workload composition and arrival rates derived from real-world traces in the Azure Public Dataset~\cite{patel2024splitwise}.}
    \label{fig:temporal}
\end{figure}

\textbf{Spatial trade-offs.} As illustrated in~\autoref{fig:spatial}, the optimal model deployment configuration (i.e., resource allocation and parallelism strategy) varies significantly across workload types. Compute-intensive workloads with short outputs benefit from data parallelism to maximize computational throughput, whereas memory-intensive workloads with long outputs favor model parallelism to distribute memory requirements. Existing systems that employ \textbf{homogeneous} deployment configurations~\cite{yu2022orca,sun2024llumnix} across all model replicas fail to account for the diverse parallelism and resource demands from different workloads, inevitably sacrificing efficiency.

\textbf{Temporal trade-offs.} As illustrated in~\autoref{fig:temporal}, workload composition and arrival rates exhibit significant temporal variations. For instance, tasks with short outputs may predominate during business hours with high arrival rates, while tasks with long outputs increase in the evening with lower traffic, leading to shifting resource demands across different periods~\cite{wang2024burstgpt,stojkovic2024dynamollm}. Existing systems that employ \textbf{static} deployment configurations~\cite{kwon2023efficient,li2023alpaserve} fail to accommodate these time-varying demands in resource allocation and parallelism strategies, resulting in suboptimal resource utilization and performance degradation.

Motivated by these, this work develops \sys, an efficient LLM serving system that tackles the workload heterogeneity on both the \textbf{spatial} and \textbf{temporal} dimensions.
Essentially, \sys makes two innovations as follows.

\noindent \textbf{\underline{I1:} Workload-aware scheduling to address spatial heterogeneity with heterogeneous model deployment.} 
To handle the mixed workload composition, \sys deploys model replicas with non-unique resources and strategies, offering the possibility to distinguish the serving of different workload types spatially.
However, this heterogeneous approach adds complexity in determining the optimal model deployment, and needs to meticulously assign the incoming requests across the heterogeneous replicas for workload balance. 
To achieve so, we formulate the search for heterogeneous model deployments as a \textit{constrained optimization problem} and propose a \textit{two-level workload-aware scheduling} algorithm.
This approach co-optimizes model deployment with workload assignment, ensures that each model replica is configured and utilized in the best possible way to meet the specific demands of different workloads.

\noindent \textbf{\underline{I2:} Workload-adaptive switching to address temporal heterogeneity with flexible model deployment.}  
In response to the dynamicity in workload composition and arrival rates, \sys enables adaptively switching the model deployment (as well as the request assignment) to gain flexibility on the temporal dimension. 
Two major efforts are made to do this. 
Firstly, we devise a \textit{fine-grained time-series forecasting}, which accurately predicts how the workloads change in the next time interval.
Secondly, to expedite the switching of model deployment, we introduce a \textit{ad hoc model switching} method, which avoids re-loading the huge model from scratch, but leverages the faster GPU-GPU network connections to transfer model parameters.

We conduct experiments to compare \sys with vLLM, Llumnix, and Dynamo on different real-world traces using popular LLMs with up to 70B parameters. 
Empirical results show that \sys reduces the end-to-end P99 tail latency and improves system throughput by up to 2$\times$ and on average 1.5$\times$ compared to state-of-the-art LLM serving systems.

\section{Background}
\label{sec:preliminary}

\textbf{Workload heterogeneity.} LLMs are designed to support a diverse range of applications, and these different inference workloads exhibit heterogeneity in terms of input and output sequence lengths~\cite{naveed2023comprehensive,hadi2024large}. For example, chat, information extraction, and document summarization tasks, as well as requests from BurstGPT and MMLU, typically have short output lengths~\cite{wang2024burstgpt,hendrycks2020measuring,patel2024splitwise}. Conversely, code/content generation, transformation, and reasoning tasks, along with requests from ShareGPT and WildChat, usually have long output lengths~\cite{zheng2023lmsys,zhao2024wildchat,gao2024omni,jain2024livecodebench}.

\noindent \textbf{Phases of LLM inference.} Given an input prompt, LLM inference consists of two phases: The prefill phase processes the prompt to compute the key-value (KV) cache and generates the first token in a single step, and the decoding phase takes the last generated token and KV cache as inputs to generate subsequent tokens~\cite{vaswani2017attention}. Unlike the prefill phase, the decoding phase generates tokens step-by-step, which makes it more memory-bandwidth-bound than the compute-intensive prefill phase~\cite{zhao2024atom}.

\noindent \textbf{Parallelisms.} To parallelize the model over multiple GPUs, there are three prevalent forms of parallelisms, which are data (i.e., model replication)~\cite{li2023alpaserve,liu2024understanding}, tensor~\cite{shoeybi2019megatron}, and pipeline parallelism~\cite{huang2019gpipe}.
Different parallelisms come with certain trade-offs. 
% Concretely, data parallelism enables parallel processing of requests but introduces additional memory overhead. Tensor and pipeline parallelism distribute model computations across devices to manage memory usage more efficiently but incur communication overhead.
Numerous studies~\cite{li2023alpaserve,zhong2024distserve,jiang2023hexgen,miao2024spotserve} have investigated how to deduce the hybrid parallelism strategy by meticulously enumerating many possible combinations.

\noindent \textbf{Job scheduling in clusters.} 
There is also a line of research that considers the job scheduling in clusters~\cite{isard2009quincy,schwarzkopf2013omega,delimitrou2013paragon,delimitrou2014quasar}. However, our work focuses on the request scheduling for LLM serving, which has a different goal. Extended related work is provided in~\autoref{sec:relatedwork}.

\section{\sys Overview}
\label{appendix:overview}

\begin{figure}[!t]
    \centering
    \includegraphics[width=\linewidth]{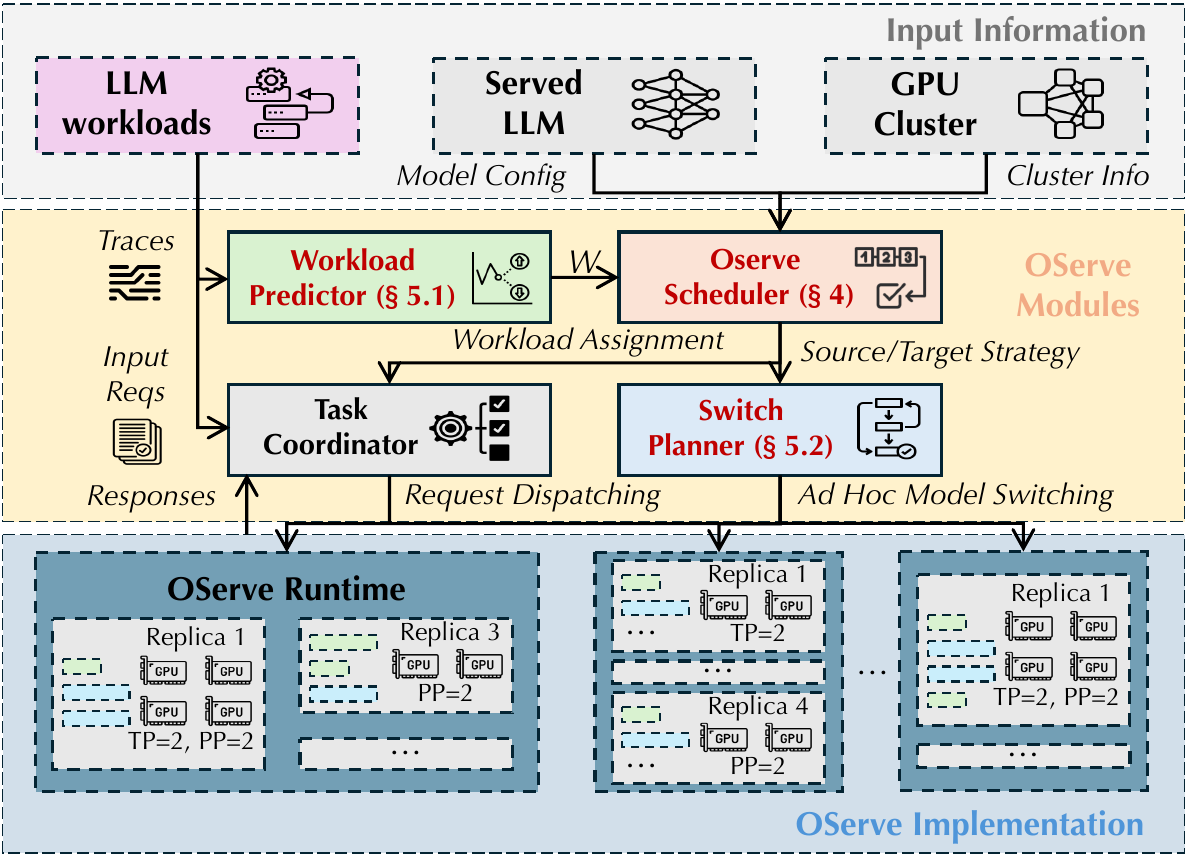}
    \caption{\sys system overview.}
    \label{fig:system overview}
\end{figure}

The architecture overview of \sys is shown in~\autoref{fig:system overview}. \sys contains three essential components:
The workload predictor (\S\ref{sec:workload predictor}), \sys scheduler (\S\ref{sec:alg}), and switch planner (\S\ref{sec:smart swicth}). The overall routine is as follows. \textbf{(\underline{1}) Workload prediction:} The workload predictor takes the LLM workloads' historical traces as input, based on which it differentiates between different workload types and predicts their corresponding request arrival rates in the next time span (in our case, one minute). The workload is considered stable during this time span given the short duration of each span~\cite{duan2024parcae}. \textbf{(\underline{2}) Strategy deduction:} The scheduler takes the cluster information, model configuration, and estimated workloads as input, formulates the flow network, and deduces the optimal serving strategy. The \sys engine then deploys the LLM and performs request dispatching based on the scheduling result. \textbf{(\underline{3}) Strategy switch:} Once the workload predictor provides the workload features for the subsequent time span, the scheduler searches for the corresponding serving strategy and outputs it to the switch planner. The switch planner utilizes the source (previous) and target (current) strategies to obtain the optimal switch plan. The engine then implements model parameter switching based on the given instructions.

\section{Workload-aware Scheduling}
\label{sec:alg}

\subsection{Scheduling Problem Statement}
\label{sec:problem formulation}

\begin{figure}[!t]
    \centering
    \includegraphics[width=\linewidth]{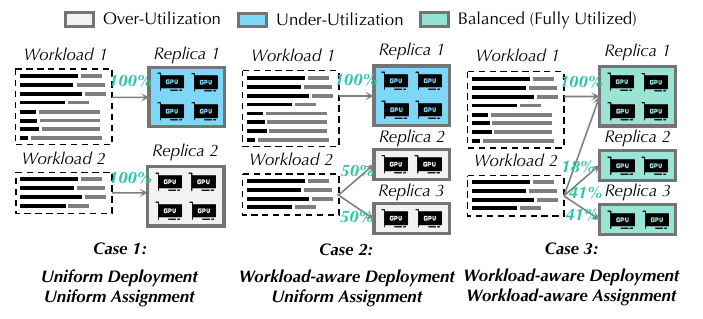}
    \caption{Example of model deployment and workload assignment.}
    \label{fig:example}
\end{figure}

To support efficient workload-aware LLM serving, the scheduling algorithm must determine two essential allocations: (\textbf{\underline{i}}) \textit{Model deployment}, which specifies the resource allocations and parallelism strategies for multiple model replicas; and (\textbf{\underline{ii}}) \textit{workload assignment}, which determines the distribution of workload types across different model replicas. We term a solution to these two components a \textit{serving strategy}. As shown in~\autoref{fig:example}, different serving strategies yield different performance results. To tackle the complexity of this problem, we propose a two-level workload-aware scheduling algorithm:

\noindent \textbf{Lower-level optimization (\S\ref{sec:first stage}):} 
Given a specific model deployment, the lower-level formulates the workload assignment problem as a directed flow network, and applies a max-flow algorithm to maximize serving performance.

\noindent \textbf{Upper-level optimization (\S\ref{sec:second stage}):}
Based on the lower-level outcome, the upper-level applies flow network-guided generation to obtain the optimal model deployment. This iteratively refines resource allocation and parallelism strategy based on system bottlenecks identified by the flow network.

Together, these two levels form a cohesive optimization loop. We illustrate the scheduling process with a simple example in \autoref{sec: simple example}.

\subsection{Lower-level Workload Assignment}
\label{sec:first stage}

The lower-level of our algorithm determines the optimal workload assignment for a specific model deployment.

\begin{figure}[!t]
    \centering
    \includegraphics[width=\linewidth]{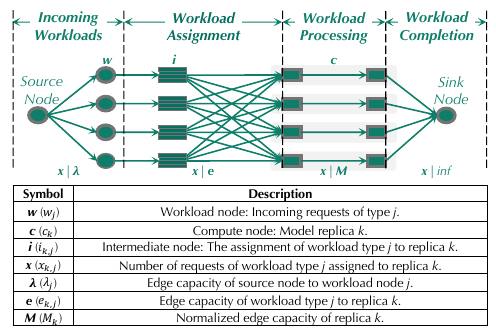}
    \caption{Illustration of the flow network. $a \mid b$ denotes that $a$ is the used capacity out of $b$.}
    \label{fig:bg}
\end{figure}

\textbf{Flow network formulation.} We construct a flow network in which the source node \(\mathcal{S}\) supplies all incoming requests and the sink node \(\mathcal{T}\) represents completed requests. Each workload type \(j\) corresponds to a workload node \(w_j\), and each model replica \(k\) is represented by two nodes, \(c_k^{in}\) and \(c_k^{out}\), where \(n_{k,j}\) denotes the processing capacity of replica \(k\) for workload type \(j\). To facilitate per-workload assignments, we introduce intermediate nodes \(i_{k,j}\). The flow network comprises four types of edges:
(\underline{\textbf{i}}) From the source \(\mathcal{S}\) to each workload node \(w_j\), with capacity $\lambda_j$ equal to the total number of incoming requests of type \(j\);
(\underline{\textbf{ii}}) from each workload node \(w_j\) through an intermediate node \(i_{k,j}\) to \(c_k^{in}\), with capacity \(e_{k,j}\) representing the maximum number of type-\(j\) requests assignable to replica \(k\);
(\underline{\textbf{iii}}) from \(c_k^{in}\) to \(c_k^{out}\), with normalized capacity \(M_k\) to accommodate mixed workloads (detailed below); and
(\underline{\textbf{iv}}) from \(c_k^{out}\) to the sink \(\mathcal{T}\), with sufficiently large capacity to ensure that all processed requests can exit the system.
Following prior work~\cite{patel2024splitwise,jaiswal2025sageserve,lin2024apex}, we perform one-time profiling (detailed in~\autoref{sec: one time profiling}) to estimate the node capacity \(n_{k,j}\) and edge capacity \(e_{k,j}\) for each replica-workload pair \((k,j)\), capturing the maximum achievable throughput per workload type based on parallelism strategies. An illustration of the flow network is provided in~\autoref{fig:bg}.

\noindent \textbf{Edge capacity normalization.} To handle mixed workloads on a single model replica, we compute the least common multiple \(M_k\) of all $n_{k,j}$ for replica $k$. By setting the capacity of $(c_k^{in}, c_k^{out})$ to $M_k$, we ensure that each type-$j$ request consumes $M_k/n_{k,j}$ units of capacity. For example, if replica $k$ can process $80$ type-$1$ requests and $50$ type-$2$ requests per unit time, we set the capacity of $(c_k^{in}, c_k^{out})$ to the least common multiple, i.e., $400$. Thus, one type-$1$ request consumes $400/80=5$ units of capacity while one type-$2$ request consumes $400/50=8$ units.

\textbf{Formulation of constraints.} We denote by $x_{k,j}$ the number of requests of workload type $j$ assigned to replica $k$, and impose the following constraints to ensure a feasible and optimal assignment:

\noindent \textit{\underline{C1:} Workload demand constraint.} This constraint ensures that the total assignment does not exceed the incoming request volume for each workload type: $\sum_{k=1}^{M} x_{k,j} \le \lambda_j,\ \forall j$. This aligns with the supply-side limitation, preventing assignment of more type-$j$ requests than are actually available.

\noindent \textit{\underline{C2:} Edge capacity constraint.} This constraint enforces per-replica capacity limits for individual workload types: $x_{k,j} \le e_{k,j},\ \forall k,j$. This reflects the replica-level, per-workload capacity derived from the profiling step, ensuring that the number of type-$j$ requests directed to replica $k$ does not exceed the configured edge capacity.

\noindent \textit{\underline{C3:} Node capacity sharing constraint.} This constraint governs the joint processing capacity when a replica handles multiple workload types simultaneously: $\sum_{j=1}^{J} \frac{x_{k,j} \cdot M_k}{n_{k,j}} \le M_k,\ \forall k$, where the least common multiple $M_k$ normalizes capacities across all workloads for replica $k$. This provides a compositional capacity limit, ensuring that the combined resource consumption of different workload types does not exceed the total processing capability of each replica.

Given the three constraints and the flow network formulation, we solve the resulting max-flow problem using the \textit{preflow-push algorithm}~\cite{cheriyan1989analysis}, which computes the optimal workload assignment $\{x_{k,j}\}$.

\subsection{Upper-Level Model Deployment}
\label{sec:second stage}

Based on the lower-level optimization, the upper-level problem determines the optimal model deployment to maximize overall serving performance.

\noindent \textbf{Problem formulation.} Consider a cluster of \(D\) GPUs with the objective of forming \(R\) model replicas (i.e., data parallelism). Let \(\{d_r\}_{r=1}^{R}\) denote the number of GPUs assigned to each replica and \(\{s_r\}_{r=1}^{R}\) the corresponding parallelism strategies (i.e., tensor and pipeline parallelism), subject to: \(\sum_{r=1}^{R} d_r = D, \ \forall r.\) For any configuration \(\{d_r, s_r\}_{r=1}^{R}\), the lower-level optimization (\S\ref{sec:first stage}) yields the maximum achievable throughput \(\Phi(\{d_r, s_r\})\). The upper-level problem thus seeks to solve: $\max_{\{d_r,s_r\}} \Phi(\{d_r,s_r\}).$ 

A brute-force approach enumerating all valid configurations is feasible for small \(D\) and \(R\), but becomes intractable as the search space grows. To address this, we propose a flow network guided generation method that iteratively searches for the optimal deployment configuration.

\noindent \textbf{Initialization.} Prior to the search process, we initialize a uniform model deployment by allocating an identical number of GPUs to each replica and employing pure tensor parallelism. The GPU count per replica is determined by the minimum memory requirement necessary to serve the model, e.g., 140 GB for a 70B model.

\begin{figure}[!t]
    \centering
    \includegraphics[width=0.8\linewidth]{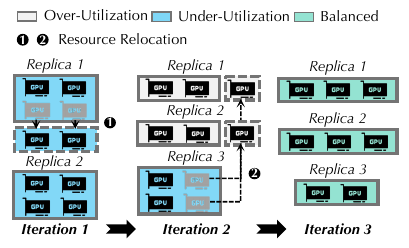}
    \caption{Example of flow network guided generation.}
    \label{fig:mut}
\end{figure}

\noindent \textbf{Flow network guided generation.}
To efficiently navigate the large search space, we leverage insights from the lower-level max-flow analysis. Specifically, after solving the max-flow problem for a given deployment, we examine the flow distribution across different replicas to identify: (\textbf{\underline{i}}) Bottleneck replicas that are fully saturated and could benefit from additional GPU resources, and (\textbf{\underline{ii}}) under-utilized replicas whose assigned GPUs are not fully exploited. These insights guide a heuristic refinement process in which GPUs are reallocated from under-utilized replicas to bottleneck replicas, iteratively improving overall throughput. Each iteration consists of three essential steps:
\begin{itemize}[topsep=5pt, leftmargin=*]
    \vspace{-0.75em}
    \item Identify bottleneck and under-utilized replicas by comparing actual node flow against capacity.
    \vspace{-0.75em}
    \item Reallocate GPUs, increasing \(d_r\) for bottleneck replicas and decreasing it for under-utilized ones.
    \vspace{-0.75em}
    \item Enumerate possible parallelization strategies \(\{s_r\}\) for the adjusted configuration, selecting the one that yields the highest throughput.
    \vspace{-0.75em}
\end{itemize}

\autoref{fig:mut} illustrates this iterative process. By continuously refining resource allocation and parallelism strategy (i.e., \(\{d_r, s_r\}\)) based on flow network feedback, the upper-level optimization converges toward an efficient model deployment that maximizes system throughput.

\noindent \textbf{Termination condition.} We terminate the iterative search process when no further improvements can be made. For instance, when the maximum achievable throughput remains unchanged for 20 iterations.

\section{Workload-adaptive Switching}

\subsection{Workload Prediction}
\label{sec:workload predictor}

\textbf{Challenges in workload prediction.} Three key characteristics define an LLM inference workload: (\underline{\textbf{i}}) Input request sequence length, (\underline{\textbf{ii}}) output request sequence length, and (\underline{\textbf{iii}}) request arrival rate. However, direct prediction of these variables is impractical due to task variability and complexity. For example, in the Azure Public Dataset~\cite{patel2024splitwise}, the input sequence length ranges from $1$ to $7999$, and the output sequence length ranges from $1$ to $5000$. Furthermore, given the dynamic and unpredictable nature of user interactions, the request arrival rate can exhibit substantial fluctuations within even brief time spans~\cite{qiao2024conserve,stojkovic2024dynamollm,wang2024towards}.

\noindent \textbf{Our approach.} Motivated by previous works~\cite{ma2018query,duan2024parcae} on \textbf{clustering} and \textbf{statistical}-based prediction, \sys employs a \textit{fine-grained time-series forecasting} approach. Rather than forecasting the exact lengths of input and output request sequence lengths and arrival rates, we focus on workload differentiation and type-specific prediction. Our method enables \sys to: (\underline{\textbf{i}}) Distinguish between different workload types, and (\underline{\textbf{ii}}) independently predict the estimated number of requests for each type over short future time spans.

\noindent \textbf{Future workload estimation.} Although it is hard to predict the concrete future workload information, we can separate the workload into different types based on historical input and output request sequence lengths, and predict each of the workload type's the arrival rate, i.e., how many requests of each type are arrived within the next time span (one minute in our work\footnote{We assume that the workload is relatively stable within each time span, which is a reasonable assumption given the short duration of each span.}).
Our approach consists two steps:

\textit{\underline{S1:}} \textit{Process historical data.} We utilize a k-means algorithm~\cite{ahmed2020k} to categorize historical data into distinct workload types based on input and output request sequence lengths, ensuring each request is assigned to a specific workload type. We then count the number of requests that arrive within each time span for each workload type. This processed historical data enables us to establish the relationship between the total number of arrived requests and the corresponding time spans. 

\textit{\underline{S2:}} \textit{LSTM prediction.} We select the Long Short-Term Memory (LSTM) model~\cite{yu2019review} as our workload predictor due to its superior ability to capture long-range temporal dependencies~\cite{chien2021slowerlstm,zhang2021recurrentlstm}. It uses a sequence length of 50, enabling it to leverage data from the previous 50 minutes to predict workload patterns for the next minute. We train it using the processed historical data mentioned in the first step.

\begin{figure}[!t]
    \centering
    \includegraphics[width=\linewidth]{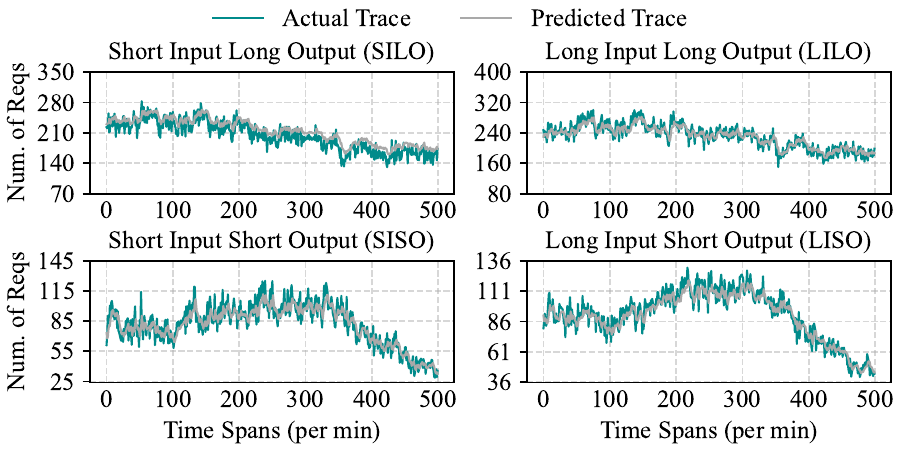}
    \caption{Prediction of the arrival rate (i.e., the number of requests per time span) for each workload type.}
    \label{fig:predi}
\end{figure}

\noindent \textbf{Predictor performance.} We train an LSTM model using the two-step method with two weeks of real-world traces from the Azure Public Dataset, with 90\% of the data as the training set and 10\% as the test set. As shown in \autoref{fig:predi}, the LSTM model efficiently and accurately captures the trends of future inference workloads. The workload prediction process takes less than 30 ms with an average Relative Root Mean Square Error (RRMSE) of 5.045\%, which demonstrates the effectiveness of our approach.

\subsection{Ad Hoc Model Switching}
\label{sec:smart swicth}

\noindent \textbf{Challenges in model switching.} Model switching occurs when the optimal serving strategy shifts due to workload fluctuations. Due to the substantial sizes of LLMs, re-loading the model from scratch is time-consuming (taking minutes as evaluated in \S\ref{sec:expr}). Such long switching times degrade performance under highly fluctuating workloads and diminish the benefits of adopting the new serving strategy.

\begin{figure}[!t]
    \centering
    \includegraphics[width=0.8\linewidth]{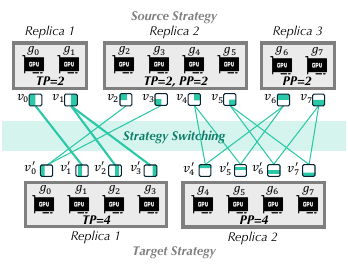}
    \caption{Example of strategy switching between eight GPUs ($g_0$-$g_7$), and the model parameter shards on each GPU before ($v_0$-$v_7$) and after ($v_0'$-$v_7'$) strategy switching. $v_1$ transmits part of its model parameters to $v_2'$, since it contains the model parameters $v_2'$ required; $v_2'$ can also obtain these parameters from $v_4$ and $v_5$.}
    \label{fig:strategy switching}
\end{figure}

\noindent \textbf{Our approach.} Since all model replicas share the same parameters, \sys leverages high-speed GPU interconnects (e.g., NVLink and InfiniBand) to transfer model parameters between GPUs when switching deployments. We define the model deployments before and after switching as the \textit{source strategy} and \textit{target strategy}, and each GPU can act as both a \textit{source device} and a \textit{target device}. As illustrated in \autoref{fig:strategy switching}, each model parameter may correspond to different numbers of source and target devices depending on the source and target strategies, resulting in numerous possible communication/switch plans. Our objective is to identify the optimal switch plan that minimizes the switching cost.

\noindent \textbf{Greedy Algorithm.} We develop a \textit{greedy algorithm} to obtain the optimal switch plan. The algorithm initializes an empty plan for parameter transfers and sets up counters to track the volume of data communicated between each source-target device pair. For each target device, the algorithm iterates through all possible source devices, selects the one with the lowest existing communication load, updates the communication load by adding the required volume, and includes the selected source-target pair in the switch plan. This approach ensures that, for each model shard, the source device with the least data sent so far is always chosen, thereby optimizing overall communication efficiency.

The efficiency of this greedy algorithm can be enhanced using the heuristic that \textit{intra-machine communication should always be prioritized over inter-machine communication}. NVLink (400 GB/s) handles intra-machine communication, which is typically faster than inter-machine communication using InfiniBand (IB) or RoCE (10-200 GB/s). By applying this heuristic, the iteration process is divided into two phases: First iterating among intra-machine source devices, and then, if no intra-machine sources are available, iterating among inter-machine sources. This heuristic prunes unnecessary inter-machine transmission searches for each target device, thereby accelerating algorithm convergence.

Although the heuristic-based greedy algorithm does not guarantee an optimal switch plan in all cases, it is practical for real-world GPU clusters and effectively balances communication load across devices. The detailed procedure and pseudocode are provided in \autoref{appendix:greedy algorithm}.

\noindent \textbf{KV cache transmission.} For KV cache transmission: At switch time, (\underline{\textbf{i}}) requests with short-sequence KV blocks are drained on the source deployment, whereas (\underline{\textbf{ii}}) long-sequence KV blocks are migrated using the same greedy algorithm as for model parameters, leveraging fast communication links for efficient transmission. To prevent allocation stalls, we pre-allocate fixed-size KV buffers on target GPUs sized to the required KV capacity (optionally +10–20\% headroom for fragmentation), and migrate KV in batched, layer-aligned chunks. Once a request drains or completes migration, the source buffers are reclaimed.

\section{Experimental Evaluation}
\label{sec:expr}

\subsection{Experimental Setup}
\label{sec:expr_setup}

\textbf{Environments.} Our experiments are conducted on four GPU servers equipped with \texttt{8$\times$NVIDIA H100-80GB} GPUs. Within each server, the GPUs are connected via NVLink with a bandwidth of 400GB/s, and the servers are connected via InifiBand with a bandwidth of 200GB/s.

\noindent \textbf{Baselines.} To understand the system efficiency of \sys and each part of our system design, we compare it with state-of-the-art LLM serving systems: (\textbf{\underline{i}}) vLLM (static)~\cite{kwon2023efficient}: Using vLLM to serve the given LLM with a static parallel configuration. (\textbf{\underline{ii}}) vLLM (reload): Enabling vLLM with \switch (\S\ref{sec:smart swicth}) to serve with adjusted model deployments to adapt to different inference workloads. (\textbf{\underline{iii}}) Llumnix~\cite{sun2024llumnix}: Continuously rescheduling and dynamically migrating requests across instances to handle workload fluctuations. (\textbf{\underline{iv}}) Dynamo+vLLM~\cite{nvidia_dynamo_github_2025}: Nvidia's distributed inference framework that dynamically rebalances GPUs and routes \seqsplit{requests/KV-cache} to reduce recompute and queuing. To ensure a fair comparison, we optimize the model deployment for each baseline and report the best results.
% \begin{itemize}[topsep=5pt, leftmargin=*]
%     \vspace{-0.75em}
%     \item vLLM (static)~\cite{kwon2023efficient}: using vLLM to serve the given LLM with a static parallel configuration. To ensure a fair comparison, we optimize the model deployment for vLLM and report the best results.
%     \vspace{-0.75em}
%     \item vLLM (reload): enabling vLLM with \switch (\S\ref{sec:smart swicth}) to serve with adjusted model deployments to adapt to different inference workloads.
%     \vspace{-0.75em}
%     \item Llumnix~\cite{sun2024llumnix}: It continuously reschedules and dynamically migrates requests across instances to handle workload fluctuations. We optimize the model deployment for Llumnix and report the best results.
%     \vspace{-0.75em}
%     \item 
%     Dynamo+vLLM~\cite{nvidia_dynamo_github_2025}: Nvidia's distributed inference framework that dynamically rebalances GPUs and routes \seqsplit{requests/KV-cache} to reduce recompute and queuing. We also report the best performance.
%     \vspace{-0.75em}
% \end{itemize}

\noindent \textbf{Models.} Similar to prior works~\cite{zhong2024distserve,kwon2023efficient}, we evaluate \sys on LLMs with different scales, including OPT-30B and OPT-66B~\cite{zhang2022opt}, Llama-30B and Llama2-70B~\cite{touvron2023llama}. 
% , which are popular LLMs widely used in academia and industry.

\begin{figure}[!t]
    \centering
    % \vspace{-0.5em}
    \includegraphics[width=\linewidth]{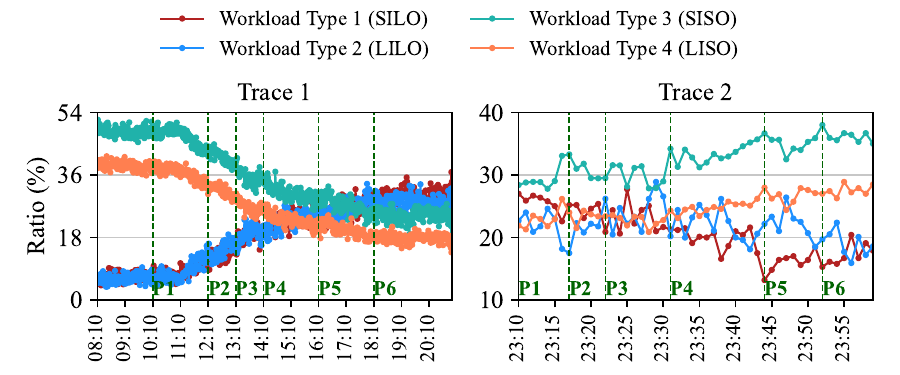}
    \caption{Workload type ratio changes in real traces.}
    \label{fig:workload variation}
\end{figure}

\noindent \textbf{LLM inference workloads.} We follow previous works~\cite{patel2024splitwise,stojkovic2024dynamollm,zhao2024blendserve} to generate workload traces from real-world datasets~\cite{azuredataset,zhao2024wildchat}. 
Two traces are sampled to evaluate \sys, as shown in \autoref{fig:workload variation}. Trace 1 (T1) represents an 800-minute period with fluctuating workloads, while trace 2 represents a 50-minute period with different fluctuation trends. In both cases, we scale the request arrival rate based on the cluster size while maintaining the workload type ratios for each minute, ensuring the cluster capacity is neither over- nor under-utilized. And we select six time spans (P1-P6) from each trace, each with a distinct workload composition, to provide a detailed performance comparison for each specific time span.

\noindent \textbf{Evaluation metrics.} We focus on a range of percentile latencies (i.e., average, P90, and P95-99) and throughput when evaluating system performance.

\begin{figure*}
    \centering
    \includegraphics[width=\linewidth]{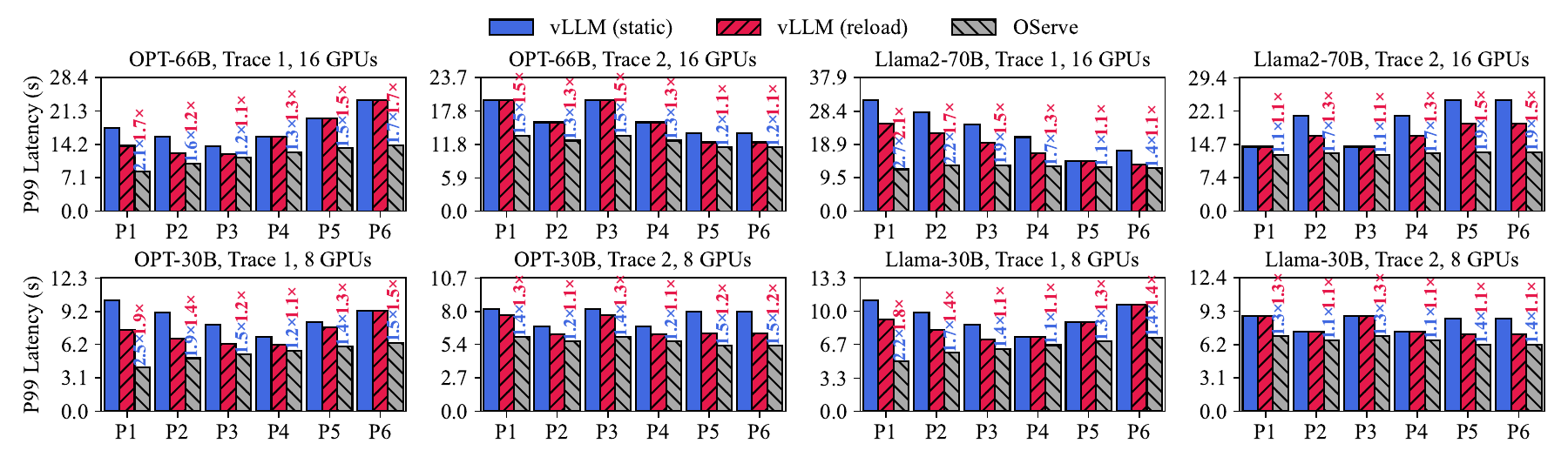}
    \caption{Latency results of \sys vs. baselines with different LLMs, GPU, and traces during different time spans (P1-P6).}
    \label{fig:ablation}
\end{figure*}

\begin{figure}[t!]
    \centering
    \includegraphics[width=\linewidth]{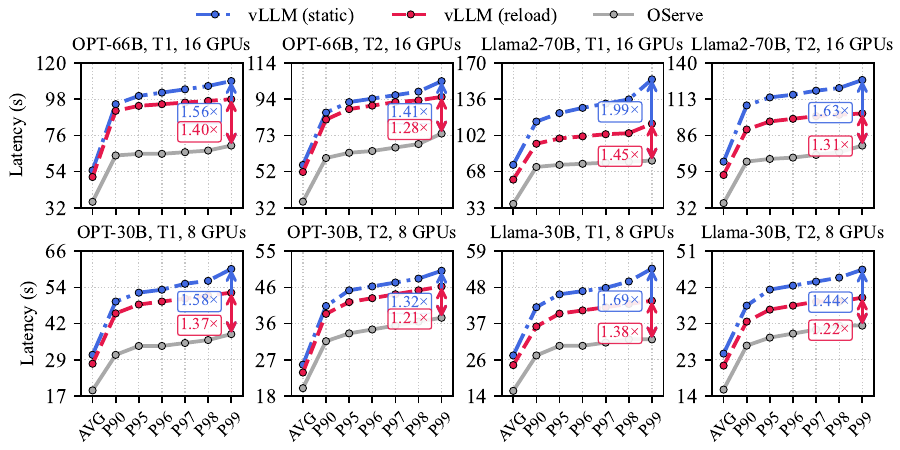}
    \caption{End-to-end latency results of \sys vs. baselines with different LLMs, GPU, and traces.}
    \label{fig:e2e}
\end{figure}

\subsection{End-to-end Performance}

\label{sec:expr_e2e}
\begin{figure}[t!]
    \centering
    \includegraphics[width=\linewidth]{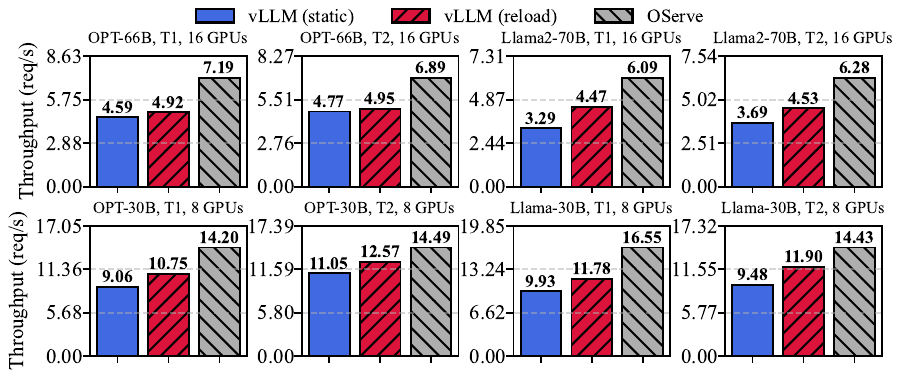}
    \caption{Throughput results of \sys vs. baselines.}
    \label{fig:thpt}
\end{figure}

\textbf{End-to-end performance comparison.} \autoref{fig:ablation}, \autoref{fig:e2e}, and \autoref{fig:thpt} demonstrate the latency and throughput performance of \sys compared with vLLM (static) and vLLM (reload) with different configurations. \sys outperforms both baselines in terms of all latency and throughput metrics. In the end-to-end experiments, \sys improves P99 latency and throughput by up to 2.0$\times$ and on average by 1.5$\times$ compared to vLLM (static), and by up to 1.5$\times$ and on average by 1.3$\times$ compared to vLLM (reload). Across each specific time span (P1–P6) from trace 1 to trace 2, the performance gains of \sys over vLLM (static) and vLLM (reload) range from 1.1$\times$ to 2.7$\times$ and from 1.1$\times$ to 1.9$\times$, depending on the workload composition.

\noindent \textbf{Temporal flexibility with flexible deployment.} \seqsplit{Compared} to vLLM (static), the key advantage of \sys lies in its ability to predict future workloads and adjust serving strategies with minimal switching costs. \sys with temporal flexibility is able to modify model deployments and workload assignments in response to varying workload compositions. In comparison, a static model deployment inevitably leads to suboptimal performance during significant workload fluctuations. For example, at P1 and P6 in trace 1, the optimal model deployments for vLLM are (DP=8, TP=2)\footnote{Represents DP and TP degrees of 8 and 2.} and (DP=2, TP=8), as the workload at P1 benefits from greater data parallelism, while P6 benefits more from increased tensor parallelism. In this scenario, the static setup of vLLM (static) leads to performance degradation of up to 2.7$\times$ compared to \sys. By predicting workload changes and dynamically adjusting its model deployment, \sys maintains optimal performance.

\noindent \textbf{Spatial flexibility with heterogeneous deployment.} \seqsplit{Compared} to vLLM (reload), the key advantage of \sys lies in its ability to flexibly allocate resources, adjust parallelism strategies, and strategically assign workloads to the most suitable model replicas. \sys with spatial flexibility is able to implement a heterogeneous model deployment, ensuring that different workload types are directed to replicas that best match their resource needs. For example, at P5 in trace 2, vLLM (reload) uses a model deployment of (DP=4, TP=2, PP=2). In contrast, \sys adopts a heterogeneous model deployment, deploying four model replicas (i.e., DP=4) with different configurations: (TP=4, PP=2), (TP=2, PP=2), (TP=2, PP=1), and (TP=2, PP=1). Each workload type is routed to the most suitable replica: Types 1 and 2, benefiting from model parallelism, are directed to replicas with more resources (i.e., TP=4, PP=2), while types 3 and 4, benefiting from data parallelism, are assigned to more replicas with fewer resources (i.e., TP=2, PP=1). In this scenario, the uniform model deployment and workload assignment in vLLM (reload) result in performance degradation of up to 1.5$\times$ compared to \sys.

\noindent \textbf{Comparison with Llumnix and Dynamo+vLLM.} We compare \sys with Llumnix, which integrates dynamic request migration, and Dynamo+vLLM, which autoscales prefill/decoding workers with KV-aware scheduling. As shown in \autoref{fig:llumnix} and \autoref{fig:dynamo} in~\autoref{appendix:baseline comparison}, when serving Llama-30B and Llama2-70B on 8–16 GPUs across multiple traces, \sys outperforms Llumnix by 1.32–1.51$\times$ in P99 latency and throughput, and improves end-to-end performance over Dynamo+vLLM by 12–20\%. Both baselines fail to account for the impact of model deployments (i.e., resource allocations and parallelism strategies) on LLM serving across various workload types: Llumnix does not consider different deployment configurations, while Dynamo fixes per-worker parallelism, overlooking the parallelism–workload interaction. In contrast, \sys co-optimizes model deployment with request scheduling, leading to significant performance improvements. Detailed experimental results and are provided in \autoref{appendix:baseline comparison}.
 
\begin{figure}[!t]
    \centering
    \includegraphics[width=\linewidth]{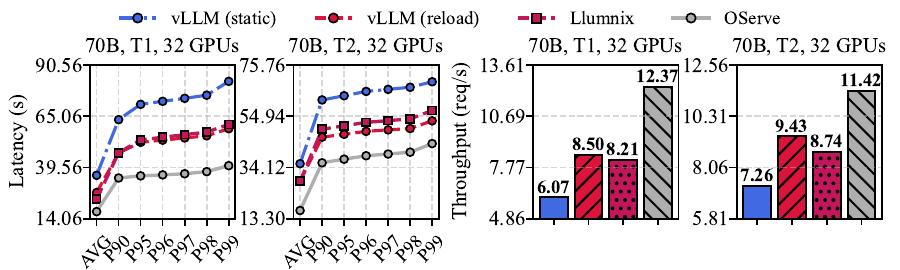}
    \caption{Results of \sys vs. baselines on 32 GPUs.}
    \label{fig:32gpus}
\end{figure}

\noindent \textbf{Experiments on a 32-GPU Cluster.}
We further evaluate \sys against the baselines on a 32-GPU cluster. As shown in~\autoref{fig:32gpus}, when serving the Llama2-70B model with 32 GPUs across different traces, \sys consistently outperforms all baselines, achieving up to a 1.9$\times$ performance improvement and demonstrating strong scalability.

\subsection{Case and Ablation Studies} 
\label{sec:expr_case_studies}
 
\begin{figure}
    \centering
    \includegraphics[width=\linewidth]{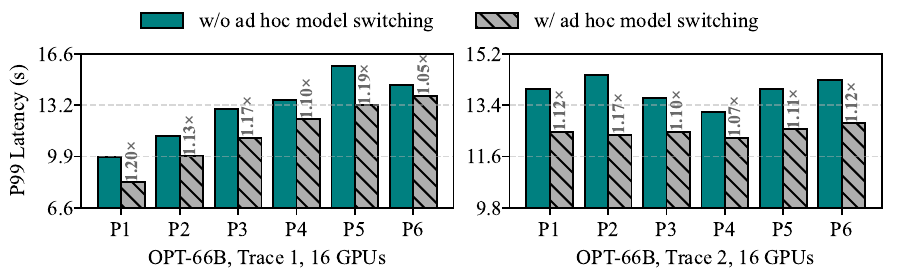}
    \caption{Switching cost impact on P99 latency.}
    \label{fig:switch}
\end{figure}

\noindent \textbf{Switching cost impact on serving latency.} The switching cost can be significant with frequently fluctuating workloads. For example, in trace 2, the minimum switching interval is 5 minutes, while reloading a model takes over 50 seconds, increasing the system's average inference latency by approximately 17\%. However, by using the \switch technique described in \S\ref{sec:smart swicth}, we minimize the switching cost to around 10 seconds for any case, significantly reducing overhead and improving \sys's adaptability to fluctuating workloads. As shown in \autoref{fig:switch}, enabling \switch reduces the system's P99 latency by up to 17\% and by an average of 12\% compared to naive model reloading. Note that the impact of \switch is more significant in workloads with higher fluctuations, requiring more frequent changes in parallel configurations.

\textbf{Comparison with other workload prediction methods.} To evaluate our LSTM-based workload predictor (\S\ref{sec:workload predictor}), we compare it against two baselines on OPT-30B with 8 GPUs using Trace 1. The first baseline is a Moving Average (MA) predictor, which increases RRMSE to 43.375\% and reduces throughput from 14.2 req/s to 10.1 req/s (41\% degradation). The second baseline uses the same LSTM architecture but predicts total workload without type decomposition; this approach fails to converge during training due to high variance and unstable temporal patterns in the aggregated signal, yielding an RRMSE of approximately 40\%. These results demonstrate that both simpler predictors and the absence of type decomposition fail to capture temporal dependencies and workload variations, validating our type-specific LSTM-based forecasting approach.

\textbf{Discussion on potential prediction errors.} Prediction errors are inevitable in real-world autoscaling scenarios~\cite{pan2023magicscaler}. To mitigate their impact, \sys employs fine-grained prediction intervals (1-minute) combined with fast switching mechanisms (\S\ref{sec:smart swicth}). Even under inaccurate predictions or unseen workload patterns, the system can quickly re-optimize deployment in the next interval, preventing long-term performance degradation.

\begin{figure}[t!]
    \centering
    \includegraphics[width=\linewidth]{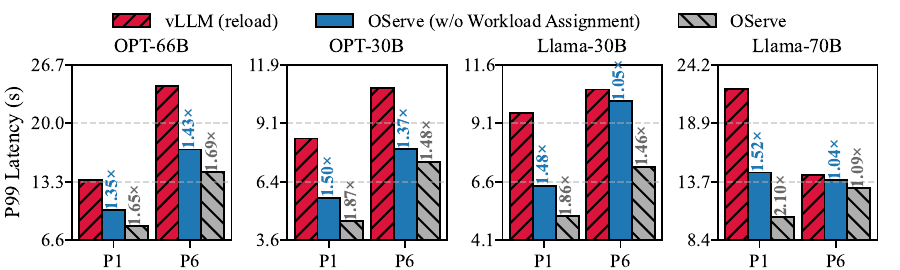}
    \caption{P99 latency of different models on two time spans (P1 and P6 in trace 1) with different \sys components.}
    \label{fig:bk}
\end{figure}

\label{sec:expr_ablation}
\textbf{Ablation study.} \autoref{fig:bk} shows the P99 latency of different models on two time spans (P1 and P6 in trace 1) with different \sys components. We start from vLLM (reload) that corresponds to optimal serving strategy of vLLM and enables each system optimization one by one. 
(\underline{\textbf{i}}) By enabling the heterogeneous model deployment, the P99 latency improves an average of 34\% and a maximum of 52\% across all cases. This heterogeneous approach allows \sys to adapt flexibly to different workload compositions by allocating varying resources to model replicas. For example, at P1 for OPT-66B, vLLM’s optimal model deployment is (DP=4, TP=2, PP=2), while \sys deploys five model replicas with configurations: (TP=3, PP=2), (TP=2, PP=2), and three with (TP=2). (\underline{\textbf{ii}}) Further enabling optimal workload assignment in \sys achieves an average improvement of 64\% and a maximum of 109\% in P99 latency. This optimization leverages heterogeneous model deployment by routing workloads to replicas that best match their resource requirements, thereby maximizing the utilization of the cluster's capabilities. For example, in the OPT-66B and P1 case, 100\% of workloads 1 and 2 are routed to the replica with (TP=3, PP=2), 88\% of workload 4 goes to the replica with (TP=2, PP=2), and the remaining 12\% of workload 4 and 100\% of workload 3 are handled by replicas with (TP=2).

% Additionally, we present a case study of the model deployments of \sys over different time spans in \autoref{appendix:casestudy}, along with TTFT (Time-To-First-Token) and TBT (Time-Between-Token) results demonstrating further improvements in key serving metrics in \autoref{appendix:ttft and tbt}.

\textbf{Sensitivity to spatial and temporal heterogeneity.}
We evaluate \sys across a spectrum of spatial and temporal heterogeneity levels on Llama2-70B with 16 GPUs.
For spatial heterogeneity, we construct five workload composition levels (S1--S5) from Azure traces with increasing skew, using the coefficient of variation (CV) of the four workload-type proportions as the indicator. 
\autoref{tab:spatial_sensitivity} shows that \sys's speedup over vLLM (static) increases from 1.14$\times$ under near-uniform workloads (CV=0.112) to 2.66$\times$ under highly skewed compositions (CV=0.688). 
For temporal heterogeneity, we construct four traces (T1--T4) with progressively shifting workload compositions across consecutive time spans, measured by the average per-type CV. 
\autoref{tab:temporal_sensitivity} shows that \sys's average speedup increases from 1.23$\times$ to 1.98$\times$ as temporal heterogeneity intensifies. 
These results demonstrate that workload-aware model deployment and assignment become increasingly effective as heterogeneity grows, while static baselines suffer from increasing mismatch to workload dynamics.

\begin{table}[t]
\centering
\small
\setlength{\tabcolsep}{4pt}
\caption{Sensitivity to spatial heterogeneity on Llama2-70B with 16 GPUs. CV is computed over the four workload-type proportions. Speedup is measured over vLLM (static).}
\label{tab:spatial_sensitivity}
\begin{tabular}{c | c | c | c}
\hline
Level & CV & SILO/LILO/SISO/LISO (\%) & Speedup \\
\hline
S1 & 0.112 & 26.3/26.2/27.3/20.2 & 1.14$\times$ \\
\hline
S2 & 0.186 & 30.0/28.0/24.2/17.8 & 1.41$\times$ \\
\hline
S3 & 0.275 & 18.8/18.9/35.5/26.8 & 1.89$\times$ \\
\hline
S4 & 0.472 & 13.2/14.2/41.1/31.5 & 2.15$\times$ \\
\hline
S5 & 0.688 & 8.2/8.2/47.1/36.6 & 2.66$\times$ \\
\hline
\end{tabular}
\end{table}

\begin{table}[t]
\centering
\small
\setlength{\tabcolsep}{5pt}
\caption{Sensitivity to temporal heterogeneity on Llama2-70B with 16 GPUs. CV is the average per-type CV across consecutive time spans. Average speedup is measured over vLLM (static).}
\label{tab:temporal_sensitivity}
\begin{tabular}{c | c | c | c}
\hline
Level & Workload Trace & CV & Avg. Speedup \\
\hline
T1 & S1$\rightarrow$S2$\rightarrow$S1 & 0.052 & 1.23$\times$ \\
\hline
T2 & S1$\rightarrow$S3$\rightarrow$S2 & 0.172 & 1.48$\times$ \\
\hline
T3 & S2$\rightarrow$S4$\rightarrow$S3 & 0.263 & 1.82$\times$ \\
\hline
T4 & S1$\rightarrow$S4$\rightarrow$S5 & 0.348 & 1.98$\times$ \\
\hline
\end{tabular}
\end{table}

Additionally, we present a case study of the model deployments of \sys over different time spans in \autoref{appendix:casestudy}, along with TTFT and TBT results demonstrating further improvements in key serving metrics in \autoref{appendix:ttft and tbt}. We also evaluate \sys on homogeneous workloads in~\autoref{appendix:homo} and discuss its extensibility to emerging parallelism strategies in~\autoref{appendix:discussion}.

\begin{figure}[t!]
    \centering
    \includegraphics[width=\linewidth]{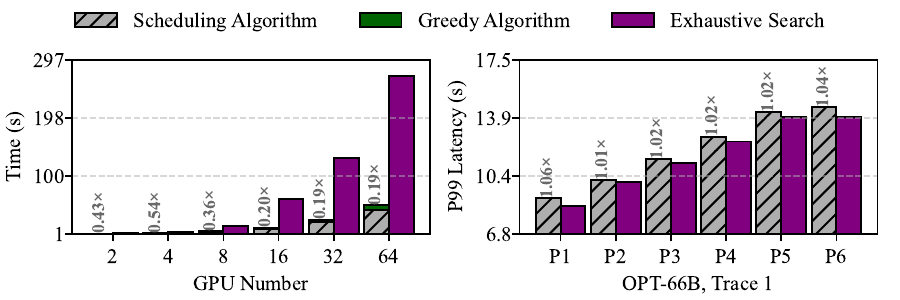}
    \caption{Algorithm scalability and impact on latency.}
    \label{fig:algorithm impact}
    \label{fig:algorithm running time}
\end{figure}

\subsection{Algorithm Efficiency}
\label{sec:expr_alg_efficiency}

\textbf{Algorithm running time.} \autoref{fig:algorithm running time} shows the execution time of our scheduling algorithm (\S\ref{sec:alg} scheduling + \S\ref{sec:smart swicth} greedy algorithm) as the number of GPUs increases. The results indicate that \sys scales well with the growing number of GPUs. Additionally, both algorithms are highly parallelizable, as different resource allocations and strategies are independent, allowing the execution time to decrease linearly with more CPU cores. Note that the scheduling time should be within the 1-minute prediction time span mentioned in \S\ref{sec:workload predictor}, as these steps need to overlap to avoid delays between scheduling and prediction. If the cluster size is extremely large and the search process takes longer than one minute, the prediction time span should be adjusted, or more CPU cores should be utilized to accelerate the computation.

\noindent \textbf{Algorithm Optimality.} To evaluate the optimality of our heuristic search, we compare the scheduling results of our heuristic design in \S\ref{sec:second stage} with those from exhaustive search, treated as the \textit{optimal} baseline. As shown in \autoref{fig:algorithm running time}, on a 16-GPU cluster, exhaustive search takes around 50s due to the large number of possible resource allocations and parallel strategies, while our heuristic method completes within 12s. Additionally, we benchmarked the P99 latencies of serving OPT-66B using scheduling results obtained through heuristic and exhaustive search. As shown in \autoref{fig:algorithm impact}, the P99 latency gap between the two methods is within 6\%, demonstrating the effectiveness of our heuristic approach.

\vspace{-1em}
\section{Conclusion}
\vspace{-0.5em}
We propose \sys, an LLM serving system that adapts to workload heterogeneity through runtime rescheduling and model deployment switching. It mitigates \textbf{temporal heterogeneity} within workload dynamics via predictive runtime model deployment switching, and addresses \textbf{spatial heterogeneity} in workload distribution through heterogeneous model deployment and adaptive workload assignment. Results demonstrate that \sys improves performance by up to 2$\times$ (average: 1.5$\times$) compared to existing systems.

% \section*{Impact Statement}
% This paper presents work whose goal is to advance the field of Machine
% Learning. There are many potential societal consequences of our work, none
% which we feel must be specifically highlighted here.

\bibliography{example_paper}
\bibliographystyle{icml2026}

%%%%%%%%%%%%%%%%%%%%%%%%%%%%%%%%%%%%%%%%%%%%%%%%%%%%%%%%%%%%%%%%%%%%%%%%%%%%%%%
%%%%%%%%%%%%%%%%%%%%%%%%%%%%%%%%%%%%%%%%%%%%%%%%%%%%%%%%%%%%%%%%%%%%%%%%%%%%%%%
% APPENDIX
%%%%%%%%%%%%%%%%%%%%%%%%%%%%%%%%%%%%%%%%%%%%%%%%%%%%%%%%%%%%%%%%%%%%%%%%%%%%%%%
%%%%%%%%%%%%%%%%%%%%%%%%%%%%%%%%%%%%%%%%%%%%%%%%%%%%%%%%%%%%%%%%%%%%%%%%%%%%%%%
\newpage
\appendix
\onecolumn

\section{Extended Related Work}
\label{sec:relatedwork}
\noindent \textbf{LLM inference serving.} There are plenty of recent researches focused on optimizing LLM inference and serving ~\cite{li2023alpaserve,kwon2023efficient,agrawal2024taming,liu2023deja,wu2023fast,zhou2022pets,yu2022orca,jiang2023hexgen,contributors2023lmdeploy,zhang2025efficient,jiang2025cascadia,wang2025thinking,yan2025fsa,zhangsurvey}. 
% Among them, 
% Orca~\cite{yu2022orca} introduces continuous batching to improve inference throughput.
AlpaServe~\cite{li2023alpaserve} adopts model parallelism to optimize LLM serving performance.
SARATHI~\cite{agrawal2024taming} introduces a chunked-prefill approach and piggybacks decoding requests to improve hardware utilization. 
% Deja Vu~\cite{liu2023deja} predicts contextual sparsity on-the-fly and uses an asynchronous and hardware-aware implementation to enhance LLM inference.
Splitwise~\cite{patel2024splitwise} splits the prefill and decoding phases onto separate machines to optimize hardware utilization. 
% TetriInfer~\cite{hu2024inference} partitions prompts into fixed-size chunks and adopts a two-level scheduling algorithm to improve the performance of disaggregated inference.
SpotServe~\cite{miao2024spotserve} supports LLM inference using preemptible instances for improving cost efficiency.
HexGen~\cite{jiang2023hexgen} proposes asymmetric parallelism and an advanced scheduling algorithm to deploy generative inference in heterogeneous environments.
We plan to explore workload-aware LLM inference with disaggregated inference architecture and heterogeneous GPU environments in future work.

\vspace{0.25em}
\noindent \textbf{Request scheduling in model inference.} Numerous systems have been developed to optimize request scheduling for model inference serving~\cite{deepspeed,gujarati2020serving,li2023alpaserve,han2022microsecond,zhang2023shepherd,sun2024llumnix,patel2024splitwise,kossmann2024gpu,peng2025hexgen}. Among them, 
% AlpaServe~\cite{li2023alpaserve} proposes a load balancing mechanism that dispatching requests according to the queue length of each individual model replica.
DeepSpeed-MII~\cite{deepspeed} uses a round-robin dispatching policy to arrange multi-instance LLM serving.
% Splitwise~\cite{patel2024splitwise} uses Join the Shortest Queue (JSQ)
% scheduling to assign a prompt and a token machine to each request.
Llumnix~\cite{sun2024llumnix} integrates dynamic request migration to ensure high throughput and low latency LLM serving, provides SLO for prioritized requests, and auto-scales instances for resource efficiency with a unified load-aware dynamic scheduling policy. \cite{kossmann2024gpu} prioritizes request according to their anticipated memory demand and the current system load, quantifies the load of each server, and routes request to server with the lowest load.
Differently, \sys distinguishes requests by type, co-optimizes request scheduling with resource allocations and parallel strategies of model replicas, and directs requests to the most suitable replicas to maximize resource utilization.

\vspace{0.25em}
\noindent \textbf{Job scheduling in clusters.} 
There is also a line of research that considers the job scheduling in clusters~\cite{isard2009quincy,schwarzkopf2013omega,delimitrou2013paragon,delimitrou2014quasar,jiang20202d,xiong2024revisiting}. However, our work focuses on the request scheduling for LLM serving, which has a different goal.

\vspace{0.25em}
\textbf{Hybrid model parallelism.} Hybrid model parallelism combines tensor parallelism with pipeline parallelism to efficiently scale LLM training and inference beyond what either strategy achieves alone~\cite{zheng2022alpa,li2023alpaserve,miao2022galvatron,jiang2022osdp,wang2024improving,he2025efficient,yan2025areal}. Hybrid parallelism enables fitting massive models across multiple GPUs while minimizing inter-machine communication overhead and latency, which is critical for meeting real-time inference requirements.

\section{Extended Discussion}
\label{sec:discussion}

\textbf{Scheduling optimality.} 
Some of the proposed algorithms are developed based on heuristics (\S\ref{sec:alg} and \S\ref{sec:smart swicth}), which may introduce sub-optimality to the scheduling results (i.e., serving strategies and switch plans). 
However, this increases the efficiency of the algorithms by a large extent, and meanwhile the scheduling results are still effective, as evaluated in \S\ref{sec:expr_alg_efficiency}. 
Consequently, although our heuristics trade off theoretical optimality for the search efficiency, we believe they are well-suited for practical deployment in dynamic serving environments, since they ensure the algorithms generate near-optimal plans under real-time constraints. 
% Our proposed algorithms (mentioned in \S\ref{sec:alg} and \S\ref{sec:smart swicth}) do not guarantee \textbf{absolute} optimality of scheduling results, they can produce serving strategies and switch plans that are highly feasible and practical within constrained search times (e.g., the prediction time span outlined in~\S\ref{sec:workload predictor}). 
% This intentional trade-off prioritizes adaptability to dynamic workload changes over theoretical optimality, ensuring the algorithms generate near-optimal plans under real-time constraints, making them well-suited for practical deployment in dynamic serving environments.

\vspace{0.25em}
\noindent \textbf{Fault tolerance and serving with spot instances.} 
Employing heterogeneous model deployment increases the complexity to manage the model replicas, which may raise hurdles for fault tolerance.
Nevertheless, our scheduling algorithm's flexibility makes it well-suited for extension to fault-tolerant and spot instance scenarios --- upon detecting a failure or resource change, the algorithm can initiate a rerun with an updated cluster size, identify the optimal model deployment, and seamlessly adjust through model deployment switching.

% \textbf{Workload traces.} Due to the limited availability of high-quality open-source LLM inference workload traces, our experiments primarily utilized the Azure Public Dataset~\cite{patel2024splitwise}, which has been widely used as experimental traces in recent works~\cite{mei2024helix,zhao2024blendserve,stojkovic2024dynamollm}. However, we acknowledge that expanding the evaluation to include additional real-world traces from industrial settings would further enhance the impact and applicability of this work.

\vspace{0.25em}
\noindent \textbf{Extensibility to heterogeneous GPU types.} Many recent studies have focused on cost-efficient LLM serving using heterogeneous GPU resources~\cite{jiang2023hexgen,mei2024helix,miao2024spotserve,griggs2024m,patel2024splitwise,jiang2025demystifying,jiang2025hexgen,jiang2025thunderserve,tong2025parallax,jiang2026boutecostefficientllmserving}. However, adapting \sys to heterogeneous environments presents significant challenges, as the search space expands dramatically when accounting for the varying computational capabilities, memory bandwidths, and memory limits of different GPU types. We leave this extension for future exploration.

\section{Simple Example}
\label{sec: simple example}

\noindent \textbf{A simple example.} To further motivate our problem formulation, consider a cluster with 8 GPUs serving two distinct workload types. Suppose there are 100 incoming requests of workload type 1 (\(\lambda_1=100\)) and 50 incoming requests of workload type 2 (\(\lambda_2=50\)). Let \(C_{k,j}\) denote the processing rate (in requests per second) of the \(k\)-th model replica on the \(j\)-th workload, and let \(f_{k,j}\) represent the fraction of workload \(j\) assigned to replica \(k\). As shown in~\autoref{fig:example}, we consider three cases:

\textit{Case 1:} Assume a model deployment consisting of two identical replicas, each configured with \((T_{1,2}=2, P_{1,2}=2)\), and each offering rates \(C_{\sim,1}=10\) requests/s and \(C_{\sim,2}=5\) requests/s. If workload type 1 is directed entirely to the first replica (\(f_{1,1}=1\)) and workload type 2 is directed entirely to the second replica (\(f_{2,2}=1\)), the total completion time is at least \(\max(\lambda_1 f_{1,1}/C_{1,1}, \lambda_2 f_{2,2}/C_{2,2})=20\) seconds.

\textit{Case 2:} Now consider a different model configuration with three replicas, where the first replica has \((T_1=2, P_1=2)\) and the second and third replicas have \((T_{2,3}=2, P_{2,3}=1)\). Under these settings, replicas 2 and 3 each process the workloads at a rate of \(C_{\sim,1}=5\) for workload 1 and \(C_{\sim,2}=3\) for workload 2. Assigning all of workload type 1 to the first replica (\(f_{1,1}=1\)) and splitting workload type 2 evenly across replicas 2 and 3 (\(f_{2,2}=0.5\), \(f_{3,2}=0.5\)) reduces the completion time to 16.67 seconds. This improvement comes from more appropriate model deployment.

\textit{Case 3:} Using the same model deployment as in Case 2, we can further optimize workload assignment. Suppose we route all of workload type 1 and 18\% of workload type 2 to the first replica (\(f_{1,1}=1\), \(f_{1,2}=0.18\)), and distribute the remaining 82\% of workload type 2 evenly between replicas 2 and 3 (\(f_{2,2}=0.41\), \(f_{3,2}=0.41\)). By carefully balancing the fractions of requests, the completion time decreases to approximately 13.67 seconds. This improvement comes from more appropriate workload assignment.

These cases highlight the importance of jointly optimizing model deployments and workload assignments. By comparing the outcomes—20 seconds with a simple two-replica scheme, 16.67 seconds with a three-replica scheme and straightforward routing, and 13.67 seconds through careful fraction allocations—we demonstrate that more nuanced model deployment and workload assignment can substantially improve system throughput and efficiency.

\section{One-Time Profiling}
\label{sec: one time profiling}

We use a one-time profiling strategy (similar to previous works~\cite{patel2024splitwise,jaiswal2025sageserve,lin2024apex}) to evaluate model capacity under different parallelism strategies for various workload types (This approach is based on the profiling method used in Vidur~\cite{agrawal2024vidur}), which captures the following components:
\begin{itemize}[topsep=5pt, leftmargin=*]
    \item \textbf{Inference‐prefilling latency}: the latency of a single transformer layer across varying tensor‐parallel (TP) degrees and workload types.
    \item \textbf{Inference‐decoding latency}: the decoding latency of a single transformer layer under the same TP‐ and workload‐type variations.
    \item \textbf{Pipeline communication latency}: the communication latency between GPUs for various workload types.
\end{itemize}
Using these measurements, we estimate per‐request latency for any configuration by combining each layer’s TP costs (both computation and communication) with the pipeline‐parallelism (PP) communication cost.
When estimating throughput, we treat the prefill and decoding phases separately:
\begin{itemize}[topsep=5pt, leftmargin=*]
    \item \textbf{Prefill phase}: compute‐bound, with batched processing capacity determined by the sum of individual layer latencies.
    \item \textbf{Decoding phase}: memory‐bound, with batched processing capacity defined by a single latency value.
\end{itemize}
This distinction has been validated in several studies~\cite{zhong2024distserve,patel2024splitwise}.

\section{Flow Network Guided Model Deployment Generation}
\label{appendix: flownetwork}

The Flow Network Guided Generation Algorithm (\S\ref{sec:second stage}) focuses on optimizing cluster-wide model deployments by iteratively adjusting GPU allocations and model parallelization strategies. Starting with an initial distribution of GPUs across multiple replicas and a chosen strategy for each replica, the algorithm uses the lower-level flow network (\S\ref{sec:first stage}) to measure how well resources are utilized. From the generated flow assignments, it identifies overutilized replicas—those operating at full capacity—and underutilized replicas—those operating at low capacity. In response, the algorithm adaptively merges, splits, or swaps GPU resources among replicas, seeking to balance the load and improve the overall throughput.

\vspace{0.5em}
\noindent In particular, overutilized replicas may merge with one another to form more efficient configurations or receive additional GPUs from underutilized replicas. Conversely, underutilized replicas might be split to form new groups or relinquish some of their GPUs to support overutilized replicas. By repeatedly adjusting these allocations and evaluating multiple parallel strategies, the algorithm converges toward an effective configuration that improves throughput. This method ensures a more dynamic and informed approach to resource management in GPU clusters, achieving higher performance and more efficient utilization than a static or one-shot allocation strategy. We demonstrate the algorithm pseudo-code in Algorithm \ref{alg:flow-guided-final}. The detailed steps are as follow:

\vspace{0.5em}
\noindent \textbf{Initialization.} The algorithm starts with a random uniform initial model deployment (with uniform resource allocation and parallel strategies), specifying how GPUs are allocated to each model replica and which parallelization strategy each replica uses.
A record of the original configuration is stored at the beginning of each iteration to allow rollback if no improvement is found.

\vspace{0.5em}
\noindent \textbf{Lower-level flow network evaluation.} In each iteration, the algorithm invokes the FlowNetwork function $\mathcal{L}$, which takes the current allocation and strategies 
$\{d_r\}$ and $\{s_r\}$ as inputs and returns a flow assignment and an achievable throughput. Based on the flow assignment, replicas are categorized as either overutilized (those operating at full capacity) or underutilized (those operating below capacity).

\vspace{0.5em}
\noindent \textbf{Adaptive resource adjustment.}
Overutilized replicas can either \texttt{merge} with another overutilized replica, combining their GPU allocations, or \texttt{swap} GPUs with an underutilized replica to balance load.
Underutilized replicas can either be \texttt{split} into two new replicas (to potentially better match resource requirements) or \texttt{swap} GPUs with an overutilized replica. Note that these mutation operations (\texttt{merge}, \texttt{split}, \texttt{swap}) are chosen randomly in each iteration, which helps keep the search unbiased and versatile.

\vspace{0.5em}
\noindent \textbf{Strategy evaluation and reversion.}
After adjusting GPU allocations, the algorithm explores all possible parallelization strategy combinations $\{s_r'\}$ to find the one that yields the highest throughput.
If the new best strategy combination leads to a higher throughput than the current recorded best, the changes are accepted. Otherwise, the algorithm reverts to the previously recorded configuration, ensuring that detrimental adjustments are discarded.

\vspace{0.5em}
\noindent \textbf{Convergence and iteration.} This process repeats until the algorithm converges (e.g., no further improvements are found) or a maximum number of iterations is reached.
Upon termination, the final $\{d_r\}$ and $\{s_r\}$ represent a model deployment that maximizes throughput according to the given constraints and cluster conditions.

\vspace{0.5em}
\noindent In essence, the algorithm dynamically negotiates the interplay between GPU distribution and parallelization strategies, driven by insights from the flow network, to elevate the cluster’s throughput and efficiency. Additional experimental results about this algorithm are demonstrated in~\S\ref{sec:expr_alg_efficiency}. The code snippets for this algorithm are available at this \href{https://anonymous.4open.science/r/LiveServe_Documents-1F54/scheduling_algorithm_documents/README.md}{URL}.

\section{Greedy Algorithm for Ad Hoc Model Switching}
\label{appendix:greedy algorithm}

The Greedy Algorithm (\S\ref{sec:smart swicth}) is designed to efficiently determine how to transfer large-scale model parameters between GPUs in order to reconfigure model deployments. Given a set of parameters, their source devices (where they are currently stored), and target devices (where they are needed), the algorithm constructs a switch plan that specifies which source device should send each required parameter shard to which target device. By initializing zeroed communication loads and incrementally adding parameter shards to minimize the communication overhead, the algorithm ensures that each parameter shard assignment is made by choosing the source device with the lowest current data transfer volume. This method inherently balances the communication load, striving to avoid bottlenecks and latency spikes.

\vspace{0.5em}
\noindent A key enhancement to the algorithm’s efficiency lies in prioritizing intra-machine communication—those transfers that occur within the same machine and thus can leverage ultra-high-speed GPU interconnects (e.g., NVLink)—over slower inter-machine communication. By first attempting to satisfy parameter requirements using intra-machine sources, the algorithm reduces overall data transfer time and network congestion. While not guaranteed to be absolutely optimal, this greedy approach provides a highly practical solution for large-scale GPU clusters, delivering near-optimal load balancing and improved parameter switching performance. We demonstrate the algorithm pseudo-code in Algorithm \ref{alg:greedy-switch}. The detailed steps are as follow:

\begin{algorithm}
\caption{Flow Network Guided Model Deployment Generation}
\label{alg:flow-guided-final}
\KwIn{ \\
  Total GPUs $D$, number of replicas $R$, 
  parallel strategy set $\mathcal{S}$, 
  FlowNetwork function $\mathcal{L}$, 
  capacities $cap(\cdot)$.
}
\KwOut{ \\
  A model deployment $(\{d_r\}, \{s_r\})$.
}

\textbf{Initialization:}\\
A random uniform model deployment $(\{d_r\}, \{s_r\})$.

\vspace{0.5em}
\Repeat{convergence or max iterations}{
    \tcp{Record current configuration}
  $\{d^{orig}_r\} \leftarrow \{d_r\}$, $\{s^{orig}_r\} \leftarrow \{s_r$\} \\
  \tcp{Flow network optimization}
  $(flow\_assignment, throughput) \leftarrow \mathcal{L}(\{d_r\}, \{s_r\})$; \\
  \tcp{Overutilized model replicas}
  $O \leftarrow \{r \mid flow\_assignment(r) = cap(r)\}$ \\
  \tcp{Underutilized model replicas}
  $U \leftarrow \{r \mid flow\_assignment(r) < cap(r)\}$

  \ForEach{$r \in O$}{
    $op \leftarrow \text{SelectOperation}(\texttt{merge}, \texttt{swap})$ \\
    \If{op = \texttt{merge} \textbf{and} $|O| > 1$}{
      \tcp{Merge two replicas into one}
      Choose $r' \in O, r' \neq r$; \\
      $d_r \leftarrow d_r + d_{r'}$; remove $r'$.
    }
    \ElseIf{op = \texttt{swap} \textbf{and} $U \neq \emptyset$}{
      \tcp{Swap GPUs between replicas}
      Choose $u \in U$ and $\delta$; \\
      $d_r \leftarrow d_r + \delta,\ d_u \leftarrow d_u - \delta$.
    }
  }

  \ForEach{$r \in U$}{
    $op \leftarrow \text{SelectOperation}(\texttt{split}, \texttt{swap})$ \\
    \If{$op = \texttt{split}$}{
      \tcp{Split one replica into two}
      Split $d_r$ into $(d_{r_1}, d_{r_2})$; \\
      replace $r$ with $r_1,r_2$.
    }
    \ElseIf{$op = \texttt{swap}$ \textbf{and} $O \neq \emptyset$}{
      \tcp{Swap GPUs between replicas}
      Choose $o \in O$ and $\delta$; \\  
      $d_o \leftarrow d_o + \delta,\ d_r \leftarrow d_r - \delta$.
    }
  }
  
  \tcp{Find the strategy combination that maximizes throughput}
  $(s'_r, throughput') \leftarrow \arg\max_{\{s'_r\} \in \mathcal{S}^R} \text{throughput}(\mathcal{L}(\{d_r\}, \{s'_r\}))$

  \If{$throughput' > throughput$}{
    $\{s_r\} \leftarrow \{s'_r\}$ \\
    $throughput \leftarrow throughput'$
  }
  \Else{
  $\{d_r\} \leftarrow \{d^{orig}_r\}$, $\{s_r\} \leftarrow \{s^{orig}_r\}$ \\
  }
}

\Return $(\{d_r\}, \{s_r\})$
\end{algorithm}

\begin{algorithm}
\caption{Greedy Switch Plan Algorithm}
\label{alg:greedy-switch}
\KwIn{ \\
  Set of model parameters $M$; For each model shard $m \in M$: a set of source devices $S_m$ that currently hold $m$, and a set of target devices $T_m$ that require $m$; Parameter volume $V_{m,t}$ for each required pair $(m,t)$; Intra-machine bandwidth priority.
}
\KwOut{ \\ A switch plan $P$ mapping each required model shard $(m,t)$ to a source device $s$.}

\textbf{Initialization:} \\ Set $P \leftarrow \emptyset$. Set $C_{s \rightarrow t} \leftarrow 0$ for all $(s,t)$ pairs.

\vspace{0.5em}
\ForEach{$m \in M$}{
  Determine $S_m$ (source devices holding $m$). \\
  Determine $T_m$ (target devices requiring $m$).

  \ForEach{$t \in T_m$}{
    \If{$t \notin S_m$}{ \tcp{Only proceed if $t$ does not already hold $m$}
      Partition $S_m$ into $S_{m}^{(\mathrm{intra})}$ and $S_{m}^{(\mathrm{inter})}$ based on intra- or inter-machine placement relative to $t$.

      \eIf{$S_{m}^{(\mathrm{intra})} \neq \emptyset$}{
        $s^{*} \leftarrow \arg\min_{s \in S_{m}^{(\mathrm{intra})}} C_{s \rightarrow t}$
      }{
        $s^{*} \leftarrow \arg\min_{s \in S_{m}^{(\mathrm{inter})}} C_{s \rightarrow t}$
      }

      Update $C_{s^{*} \rightarrow t} \leftarrow C_{s^{*} \rightarrow t} + V_{m,t}$. \\
      Update $P \leftarrow P \cup \{(m, s^{*}, t)\}$.
    }
  }
}
\Return $P$
\end{algorithm}

\vspace{0.5em}
\noindent \textbf{Initialization.} The algorithm starts with an empty switch plan $P$
It maintains a communication load counter $C_{s \rightarrow t}$, for every source-target pair 
$(s,t)$, initially zero.

\vspace{0.5em}
\noindent \textbf{Model shard assignment.} For each model shard $m$, the algorithm identifies its source devices $S_m$ (devices that currently hold $m$) and its target devices $T_m$ (devices that need $m$). For each target device $t \in T_m$, if the target already has the model shard $m$, no transfer is needed. Otherwise, the algorithm proceeds to select a suitable source device.

\vspace{0.5em}
\noindent \textbf{Intra- and inter-machine preference.} The source devices $S_m$ are partitioned into two subsets: $S_{m}^{(\mathrm{intra})}$ for devices within the same machine as $t$ and $S_{m}^{(\mathrm{inter})}$ for devices located on other machines. The algorithm first attempts to choose a source from $S_{m}^{(\mathrm{intra})}$ to exploit higher intra-machine bandwidth. If no intra-machine source is available, it falls back to selecting from $S_{m}^{(\mathrm{inter})}$.

\vspace{0.5em}
\noindent \textbf{Greedy selection.} The source device $s^*$ is chosen by finding the one with the minimum existing communication load $C_{s \rightarrow t}$. By always choosing the currently lightest-loaded source device, the algorithm spreads out data transfers and avoids creating communication hotspots. Once $s^*$ is chosen, $C_{s^* \rightarrow t}$ is updated by adding the volume $V_{m,t}$ of the transferred parameter shard, and the mapping $(m, s^*, t)$ is added to the switch plan $P$.

\vspace{0.5em}
\noindent \textbf{Outcome.} After processing all required model shards, $P$ defines a communication plan that aims to minimize the overall data transfer cost. While not provably optimal, the algorithm is a practical solution for complex, large-scale GPU clusters, providing a near-optimal communication schedule with relatively little overhead. 

\vspace{0.5em}
\noindent This method can be seamlessly integrated into larger systems where rapid reconfiguration of model replicas is crucial, ensuring quick adaptation to workload changes and resource constraints, and leading to improved overall system throughput and responsiveness. Additional experimental results about the effect of \switch and this algorithm are demonstrated in~\S\ref{sec:expr_case_studies} and~\S\ref{sec:expr_alg_efficiency}. The code snippets for this algorithm are available at this \href{https://anonymous.4open.science/r/LiveServe_Documents-1F54/greedy_algorithm_documents/README.md}{URL}.

\section{Case Study}
\label{appendix:casestudy}

\textbf{Resource allocations and parallel strategies over time.} To illustrate how allocations and strategies of \sys change over time, we benchmarked \sys's resource allocations and parallel strategies across time spans P1–P6 from trace 1 and trace 2 while serving OPT-66B models on 16 GPUs. As shown in \autoref{tab:resource_allocation_combined_vertical}, the resource allocations and parallel strategies gradually change as workloads evolve. 

\begin{table}[!t]
\centering
\caption{Resource allocations and strategies at P1-6 for trace 1 and trace 2 while serving OPT-66B models on 16 GPUs.}
\label{tab:resource_allocation_combined_vertical}
\resizebox{0.6\linewidth}{!}{
\begin{tabular}{c|c|c|c}
\hline
\textbf{Trace} & \textbf{Span} & \textbf{DP} & \textbf{Allocations and Strategies} \\
\hline
\multirow{6}{*}{Trace 1} & P1 & 5 & (TP=3, PP=2), (TP=2, PP=2), (TP=2)$\times$3 \\
\cline{2-4}
 & P2 & 4 & (TP=4, PP=2), (TP=2, PP=2), (TP=2)$\times$2 \\
\cline{2-4}
 & P3 & 4 & (PP=4), (TP=2, PP=2)$\times$3 \\
\cline{2-4}
 & P4 & 4 & (TP=3, PP=2), (TP=2), (TP=2, PP=2)$\times$2 \\
\cline{2-4}
 & P5 & 3 & (TP=4, PP=2), (TP=2, PP=2)$\times$2 \\
\cline{2-4}
 & P6 & 3 & (TP=2, PP=2), (TP=3, PP=2)$\times$2 \\
\hline
\multirow{6}{*}{Trace 2} & P1 & 4 & (TP=3, PP=2), (TP=2), (TP=2, PP=2)$\times$2 \\
\cline{2-4}
 & P2 & 3 & (TP=4, PP=2), (TP=3, PP=2), (TP=2) \\
\cline{2-4}
 & P3 & 4 & (TP=3, PP=2), (TP=2), (TP=2, PP=2)$\times$2 \\
\cline{2-4}
 & P4 & 3 & (TP=4, PP=2), (TP=2, PP=2)$\times$2 \\
\cline{2-4}
 & P5 & 4 & (TP=4, PP=2), (TP=2, PP=2), (TP=2)$\times$2 \\
\cline{2-4}
 & P6 & 5 & (TP=4, PP=2), (PP=2), (TP=2)$\times$3 \\
\hline
\end{tabular}
}
\vspace{-1em}
\end{table}

Specifically, in trace 1, during the transition from P1 to P2, two model replicas configured as (TP=3, PP=2) and (TP=2) are merged into a single replica with the configuration (TP=4, PP=2). This adjustment is driven by an increase in the proportion of workload types 1 and 2, coupled with a decrease in types 3 and 4. Compared to previous strategies, the (TP=4, PP=2) configuration improves the system's average request processing latency by approximately 20\%, thereby enhancing model serving efficiency. As the workload trend persists, the scheduling algorithm gradually allocates more GPUs to individual model replicas (i.e., increases the MP size) and reduces the number of model replicas (i.e., decreases the DP size) to adapt to the changing workload. Similarly, although trace 2 presents a different workload fluctuation trend from P1 to P6, \sys adjusts its resource allocations and parallel strategies accordingly based on the workload changes, thereby enhancing system efficiency.

\section{Compare with Llumnix and Dynamo+vLLM}
\label{appendix:baseline comparison}

\begin{figure}[!t]
    \centering
    \includegraphics[width=0.6\linewidth]{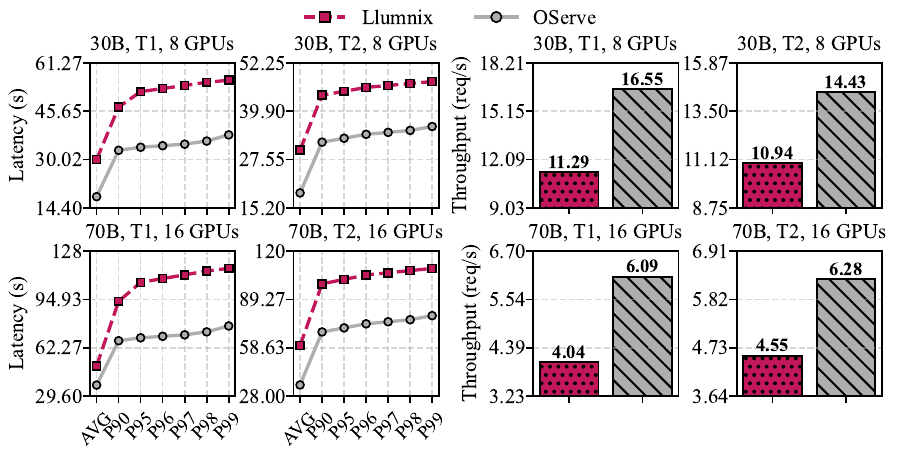}
    \vspace{-1em}
    \caption{Experimental results of \sys vs. Llumnix.}
    \vspace{-1em}
    \label{fig:llumnix}
\end{figure}

\noindent \textbf{Comparison with Llumnix.} We also compare \sys with Llumnix, a state-of-the-art inference framework that integrates dynamic request migration to handle workload fluctuations. \autoref{fig:llumnix} shows the comparison—when serving the Llama-30B and Llama2-70B models with 8 and 16 GPUs across different traces, \sys outperforms Llumnix by 1.32-1.51$\times$ in P99 latency and throughput. Llumnix does not account for the impact of different model deployments (i.e., resource allocations and parallelism strategies) on LLM serving across various workload types, limiting its ability to fully utilize system resources. In contrast, \sys co-optimizes model deployment with request scheduling, leading to significant performance improvements.

\begin{figure}[t!]
    \centering
    \includegraphics[width=0.6\linewidth]{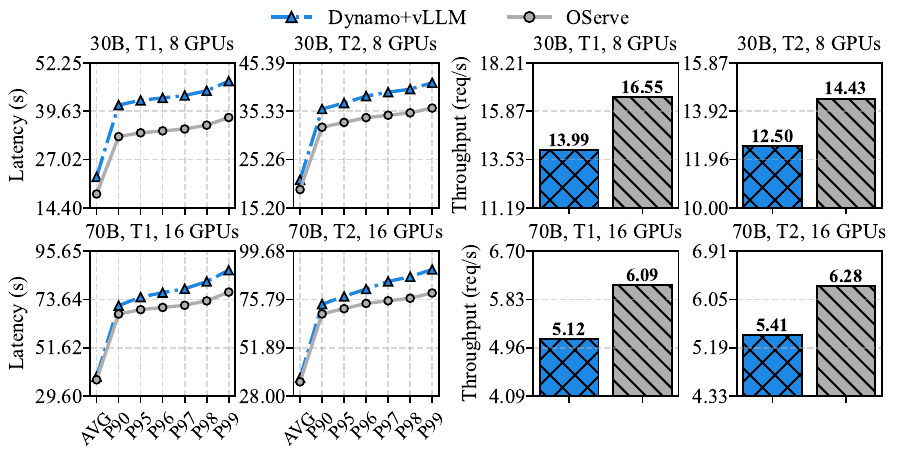}
    \vspace{-1em}
    \caption{Results of \sys vs. Dynamo+vLLM.}
    \vspace{-1em}
    \label{fig:dynamo}
\end{figure}

\noindent \textbf{Comparison with Dynamo+vLLM.} As shown in~\autoref{fig:dynamo}, we compare \sys with Dynamo+vLLM, which autoscaling prefill/decoding workers and routes requests via KV-aware scheduling. \sys improves end-to-end system performance by 12–20\% when serving Llama-30B and Llama2-70B on 8-16 GPUs across multiple traces. The gains stem from dynamically reconfiguring parallelism to match workload mix, whereas Dynamo fixes per-worker parallelism, overlooking the parallelism–workload interaction. These results highlight the benefit of \sys’s \seqsplit{co-optimization} of request scheduling and runtime parallelism.

\section{TTFT and TBT Results}
\label{appendix:ttft and tbt}

\begin{figure}[t!]
    \centering
    % \vspace{-1em}
    \includegraphics[width=0.6\linewidth]{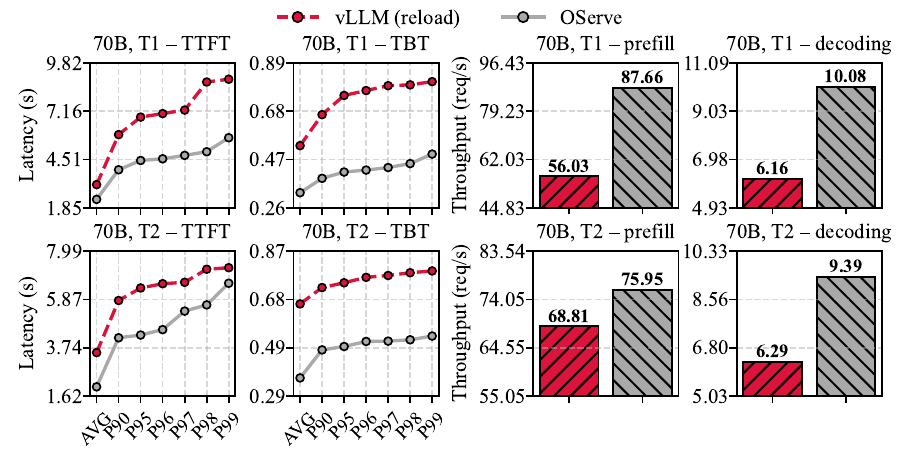}
    \vspace{-1em}
    \caption{TTFT/TBT of \sys vs. vLLM (reload).}
    \vspace{-1em}
    \label{fig:ttft_tbt}
\end{figure}

\noindent \textbf{TTFT and TBT results.} To further demonstrate the effectiveness of \sys\ in online serving scenarios, we benchmarked its TTFT (Time-To-First-Token, i.e., time to generate the first token) and TBT (Time-Between-Token, i.e., average generation time per token) against those of vLLM across various traces. As shown in \autoref{fig:ttft_tbt}, \sys\ achieves 1.1–1.6$\times$ improvements in P99 TTFT latency and 1.5–1.6$\times$ improvements in P99 TBT latency, demonstrating that, in addition to reducing end-to-end latency, \sys\ also significantly enhances other key serving metrics.

\section{Experiments on Homogeneous Workloads}
\label{appendix:homo}

\textbf{Runtime overhead on homogeneous workloads.} To verify that \sys does not introduce unnecessary overhead when workload heterogeneity is absent, we evaluate on a synthetic homogeneous trace with uniform request lengths and stable arrival rates. Under these conditions, \sys's scheduler converges to a near-uniform deployment within 2-3 iterations, achieving P99 latency within 3\% of vLLM (static) and comparable throughput. This demonstrates that \sys gracefully degrades to existing baselines when workload separability is low.

\section{Additional Discussion}
\label{appendix:discussion}

\textbf{Extensibility to emerging parallelisms.} The parallelism strategy optimization (\S\ref{sec:second stage}) is designed to be extensible. Additional parallelisms such as Expert Parallelism (EP) for MoE models and Sequence Parallelism (SP) for long-context workloads can be incorporated into the enumeration and optimization process without fundamental changes to the scheduling algorithm. This requires only profiling the new strategy's capacity characteristics and adding it to the search space.

%%%%%%%%%%%%%%%%%%%%%%%%%%%%%%%%%%%%%%%%%%%%%%%%%%%%%%%%%%%%%%%%%%%%%%%%%%%%%%%
%%%%%%%%%%%%%%%%%%%%%%%%%%%%%%%%%%%%%%%%%%%%%%%%%%%%%%%%%%%%%%%%%%%%%%%%%%%%%%%

\end{document}